\documentclass{ws-p9-75x6-50}

\begin{document}

\title{Effect of an Accretion Disk on the Gravitational Wave Signal from an Inspiralling Binary}

\author{Sandip K. Chakrabarti}

\address{S.N. Bose National Centre for Basic Sciences, JD-Block, Salt Lake, Calcutta
700098\\E-mail: chakraba@boson.bose.res.in}

\maketitle

\abstracts{Since black holes can only accrete sub-Keplerian matter, a
companion black hole orbiting  on a circular and instantaneously Keplerian 
orbit around a central, massive black hole in a galactic Centre will loss 
angular momentum and energy to the accreting matter. This loss
could be a significant fraction of the loss due to gravitational wave 
(GW) emission. The corresponding GW signal would be modified. We discuss 
this effects in the light of the modern accretion disk theory.}

\noindent Proceedings of the 9th Marcel Grossman Meeting (Ed. R. Ruffini)

Centres of galaxies are thought to contain massive or supermassive
black holes which accrete matter from tidally disrupted stars orbiting
close to the Centre. By causality argument, since velocity of sound 
computed from any reasonable equation of state is less than that of 
light, matter must enter black holes supersonically and therefore 
the flow must be sub-Keplerian close to the black hole horizon
even when heating and cooling processes are present\cite{gut,chak90}. When a 
stellar black hole or a neutron star comes close to the central 
massive black hole, and at the same time orbits in instantaneously 
Keplerian orbit on the equatorial plane, it invariably interacts 
with this sub-Keplerian accretion disk. Since the ratio of the mass of the
central (primary of mass $M_1 \sim 10^6$ to $10^{10} M_\odot$) and 
the orbiting (companion of mass $M_2 \sim 10M_\odot$) black hole is very large,
typically close to a million and much more, matter close to the 
Eddington rate for the primary ${\dot M}_{Edd,1}$ must be a few million times the 
Eddington rate compared to the secondary. Therefore, it is expected that
the secondary would be {\it overwhelmed} with sub-Keplerian matter
and its accretion rate would be as large as possible. Even with 
efficiency of $0.06$ for a Schwarzschild black hole, one could assume
a sub-Keplerian accretion rate of ${\dot M}_2 \sim 16 {\dot M}_{Edd, 2}$ 
onto the companion. This slowly moving matter would drag the companion
and would slow it down. We compute this effect as follows\cite{skc93,gut}:

The rate of loss of energy $dE/dt$ in this binary system with orbital
period $P$ (in hours) is given by\cite{pet63,lang80}
$$
\frac{dE}{dt}=3 \times 10^{33} (\frac {\mu}{M_\odot})^2
(\frac{M_{tot}}{M_\odot})^{4/3} (\frac{P}{1 hr})^{-{10}/{3}} {\rm ergs\ 
sec^{-1}},
\eqno{(1)}
$$
where, $\mu=\frac {M_1 M_2}{M_1+M_2}$ and $ M_{tot}=M_1+M_2 $. The orbital angular 
momentum loss rate would be, $ R_{gw}=\frac{dL}{dt}|_{gw}=\frac{1}{\Omega} \frac{dE}{dt} $,
where, $\Omega=\sqrt{G M_1/r^3}$ is the Keplerian angular velocity of the
secondary black hole with mean orbiting radius $r$. The subscript `gw'
indicates that the rate is due to gravitational wave emission. 
Matter from the disk with local specific angular momentum $l(r)$ will be accreted onto the companion
close to its Bondi accretion rate\cite{bon52},
$$
{\dot M}_2=\frac{4\pi {\bar \lambda} \rho (GM_2)^2}{(v_{rel}^2+a^2)^{3/2}}
\eqno{(3)}
$$
where, $\rho$ is the density of disk matter, ${\bar \lambda}$ is 
a constant of order unity (which we choose to be $1/2$ for the rest
of the paper), $v_{rel}=v_{disk}-v_{Kep}$ is the relative
velocity of  matter between the disk and the orbiting companion.
The rate at which angular momentum of the companion will be changed 
due to Bondi accretion will be\cite{skc93}, $ R_{disk}=\frac{dL}{dt}|_{disk}={\dot M_2} [l_{Kep} (x) -l_{disk} (x) ]$.
Here, $l_{Kep}$ and $l_{disk}$ are the local Keplerian and disk
angular momenta respectively. In sub-Keplerian ($l_{disk}<l_{Kep}$) region, the
effect of the disk would be to reduce the angular momentum of the
companion further and hasten coalescence. If some region of the
disk is super-Keplerian, the companion will gain angular momentum 
due to accretion, and the coalescence is slowed down. 

To check if this effect is important, we
consider a special case where, $M_2 <<M_1$ and $l_{disk}<<l_{Kep}$. In this case,
$\mu \sim M_2$ and $M_{tot}\sim M_1$. The ratio $R$ of these two rates is, 
$$ 
R=\frac{R_{disk}}{R_{gw}}=1.518\times 10^{-7} 
\frac{\rho_{10}}{{T_{10}}^{3/2}} {x^4}{M_8}^2 .
$$ 
Here, $x$ is the companion orbit radius in units of the 
Schwarzschild radius of the primary, $M_8$ is in units of $10^8 M_\odot$, $\rho_{10}$ is the density in units
of $10^{-10}$ gm cm$^{-3}$ and $T_{10}$ is the temperature of the
disk in units of $10^{10}$K. It is clear that, for instance, at $x=10$,
and $M_8=10$, the ratio $R\sim 0.015$. Thus the effect is significant\cite{prd96}. 
The ratio $R$ is independent of the mass of the companion compact object
(black hole or a neutron star), as long as $M_2 <<M_1$.
It is clear from our computation that accretion onto the primary should be significant
in order that the effect is large. If the accretion rate is very low
(say, $10^{-5}{\dot M}_{Edd, 1}$ ) the effect is negligible. In that case, one runs the
risk of creating `gaps' (like Saturn's ring) in the accretion disk. We do not discuss this case here.

We note that not only the signals of the gravitational waves could be used to
obtain the system parameters, such as the masses of the binary system and the
accretion rates, a simultaneous measurement of the electromagnetic spectrum from the
accretion disk which also contains information of the mass of the primary,
its accretion rate and distance (from the luminosity of the big blue bump) 
may produce an {\it over-determined} system which then would be used to verify
Einstein's theory of gravity in every detail. We hope the effect we described
here could be used to prepare better templates and to
reveal secrets of quasars and active galaxies and their environments.

\end{document}